\documentclass[structabstract]{aa}  
\usepackage{graphicx}
\usepackage{txfonts}
\usepackage{color}
\begin{document}
%
%\title{CXOU J170357.8-414302: a Compact Central Object inside the SNR G344.7-0.1?}
%\title{XMM-Newton and Chandra observations of the SNR G344.7-0.1 and the origin of its central object}
%\title{Exploring the origin of a central object in the SNR G344.7-0.1 with XMM-Newton and Chandra}
%\title{Infrared and X-ray emission from the SNR G344.7-0.1 and the origin of the central compact source CXOU J170357.8-414302}
%\title{A multiwavelength study of the SNR G344.7-0.1 and the central compact source CXOU J170357.8-414302}
\title{An X-ray study of the SNR G344.7-0.1 and the central object \\CXOU J170357.8-414302}
%   \subtitle{I. Overviewing the $\kappa$-mechanism}

\author{J.~A. Combi\inst{1,4}, J.F. Albacete Colombo\inst{2}, J. L\'opez-Santiago\inst{3}, G.E. Romero\inst{1,4}, E. S\'anchez-Ayaso\inst{5}, J. Mart\'{\i}\inst{5}, P.L. Luque-Escamilla\inst{6}, P.G. P\'erez-Gonz\'alez\inst{3}, A.J. Mu\~noz-Arjonilla\inst{5}, J.R. S\'anchez-Sutil\inst{5}
}

\authorrunning{Combi et~al.}

\titlerunning{X-ray study of the SNR G344.7-0.1 and CXOU J170357.8-414302}

\offprints{J.A. Combi}

   \institute{Instituto Argentino de Radioastronom\'{\i}a (CCT La Plata, CONICET), C.C.5, (1894) Villa Elisa, Buenos Aires, Argentina.\\
              \email{[jcombi:romero]@fcaglp.unlp.edu.ar}
         \and
             Centro Universitario Regional Zona Atl\'antica (CURZA). Universidad Nacional del COMAHUE, Monse\~nor Esandi y Ayacucho (8500), 
Viedma (Rio Negro), Argentina.\\
             \email{donfaca@gmail.com}
         \and
    Departamento de Astrof\'{\i}sica y Ciencias de la Atm\'osfera, Universidad Complutense de Madrid, E-28040, Madrid, Spain.\\
\email{jls@astrax.fis.ucm.es}     
\and         
Facultad de Ciencias Astron\'omicas y Geof\'{\i}sicas, Universidad Nacional de La Plata, Paseo del Bosque, B1900FWA La Plata, Argentina.
\and
         Departamento de F\'{\i}sica (EPS), Universidad de Ja\'en, Campus Las Lagunillas s/n, A3, 23071 Ja\'en, Spain\\
\email{[esayaso:jmarti:ajmunoz:jrssutil]@ujaen.es}
\and
Departamento de Ingenier\'{\i}a Mec\'anica y Minera, Escuela Polit\'ecnica Superior, Universidad de Ja\'en, Campus Las 
Lagunillas s/n, A3, 23071 Ja\'en, Spain.\\
\email{peter@ujaen.es} 
%\and
%Departamento de Astrof\'{\i}sica y Ciencias de la Atm\'osfera, Universidad Complutense de Madrid, E-28040, Madrid, Spain.\\
%\email{jls@astrax.fis.ucm.es}
             }

   \date{Received; accepted}

% \abstract{}{}{}{}{} 
% 5 {} token are mandatory
 
  \abstract
 {}
   {We report results of an X-ray study of the supernova remnant (SNR) G344.7-0.1 and the point-like X-ray source located at the
geometrical center of the SNR radio structure.}
   %, necessary to understand the SNR  evolution.
   {The morphology and spectral properties of the remnant and the central X-ray point-like source were studied using data from the XMM-Newton and Chandra satellites. Archival radio data and infrared Spitzer observations at 8 and 24 $\mu$m were used to compare and study its multi-band properties at different wavelengths.}
   {The XMM-Newton and Chandra observations reveal that the overall X-ray emission of G344.7-0.1 is extended and correlates very well with regions of bright radio and infrared emission. The X-ray spectrum is dominated by prominent atomic emission lines. These characteristics suggest that the X-ray emission originated in a thin thermal plasma, whose radiation is represented well by a plane-parallel shock plasma model (PSHOCK). Our study favors the scenario in which G344.7-0.1 is a $6 \times 10^3$ year old SNR expanding in a medium with a high density gradient and is most likely encountering a molecular cloud on the western side. In addition, we report the discovery of a soft point-like X-ray source located at the geometrical center of the radio SNR structure. The object presents some characteristics of the so-called compact central objects (CCO). However, its neutral hydrogen absorption column ($N_\mathrm{H}$) is inconsistent with that of the SNR. Coincident with the position of the source, we found infrared and optical objects with typical early-K star characteristics. The X-ray source may be a foreground star or the CCO associated with the SNR. If this latter possibility were confirmed, the point-like source would be the farthest CCO detected so far and the eighth member of the new population of isolated and weakly magnetized neutron stars.}
   {}


   \maketitle
%
%________________________________________________________________

\section{Introduction}

%%%%%%%%%%%%%%%%%%%%%%%%%%%%%%%%%%%%%%fig-1%%%%%%%%%%%%%%%%%
   \begin{figure*}
   \centering
   \includegraphics[width=18cm,angle=0]{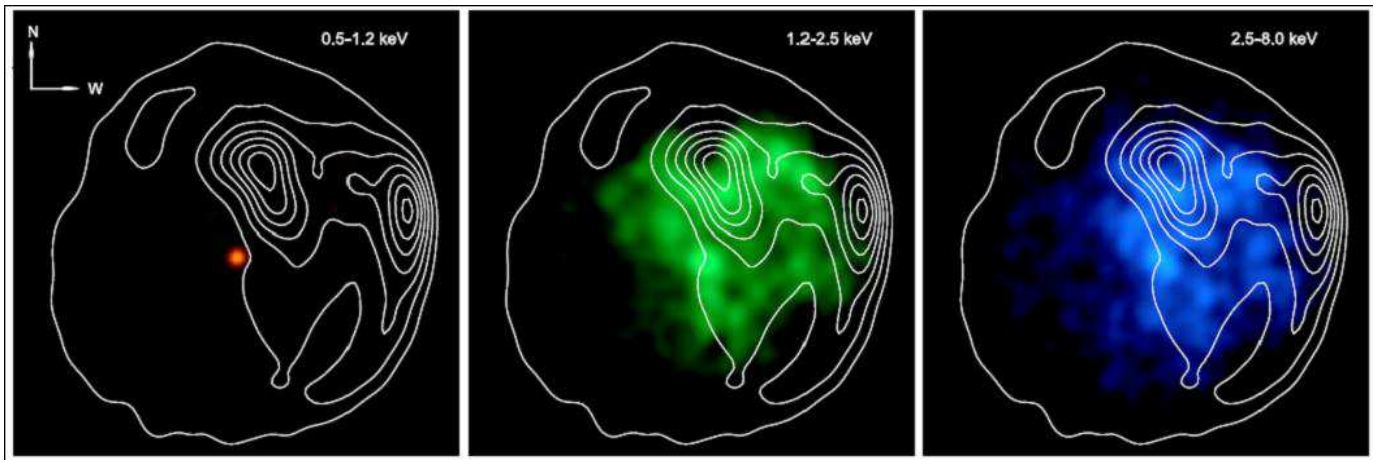}
   \caption{Chandra ACIS images, with a size of 10.4$\times$10.6 arcmin, of the SNR G344.7-0.1 
   in the three X-ray energy bands. {\bf Left panel:} soft X--rays energies (0.5-1.2 keV) in red. 
   {\bf Middle:} medium X--ray energies (1.2-2.5 keV) in green. {\bf Right:} hard X--ray energies (2.5-8 keV) 
   in blue. Smoothed images were convolved with a two-dimensional Gaussian function 
   using the {\sc aconvolve} CIAO task. Overlapping white contours are the 843
MHz radio image taken from the MOST Supernova Remnant Catalog (Whiteoak \& Green 1996).}
   \end{figure*}
%%%%%%%%%%%%%%%%%%%%%%%%%%%%%%%%%%%%%%%%%%%%%%%%%%%%%%%%%%%

The Galactic supernova remnant G344.7-0.1 was discovered in the southern sky by Caswell et al. (1975) with the
Molonglo and Parkes radio telescopes at 408 MHz and 5000 MHz, respectively. High-resolution radio imaging of the source at 1465 MHz (Dubner et al. 1993)  allowed to classify the object as a possible composite SNR. These radio observations revealed a clearly asymmetric and bright shell structure (stronger in the northwest direction) with an angular size of 8-10 arcmin. Values of radio flux densities at 408 MHz, 843 MHz, 1.47 GHz, and 
5 GHz were measured, yielding 4.7 Jy, 2.5 Jy, 1.7 Jy, and 1.3 Jy, respectively (see, Caswell et al. 1975; Dubner et al. 1993; 
Whiteoak \& Green 1996). As a result, a non-thermal spectral index of $-0.5$ (S $\propto$ $\nu^{\alpha}$) was computed for 
the source, which can be interpreted as the result of synchrotron radiation from high-energy electrons. The distance to the source is uncertain. However, Dubner et al. (1993) derived a linear diameter of $\sim$ 30 pc and a distance of about 14 kpc for the 
SNR, applying the Huang \& Thaddeeus (1985) $\Sigma$-D calibration. Throughout this work, a mean distance of 14 kpc is 
assumed.

With the advent of the ASCA satellite, which operated in the 0.2--10 keV energy range, a large number of hard X-ray sources were discovered within the Galactic plane (Sugizaki et al. 2001). Using these ASCA data, Yamauchi et al. (2005) studied the SNR G344.7-0.1 and found that the source displays  extended thermal X-ray emission with a diameter of $\sim$ 6 arcmin and that its X-ray spectrum exhibits emission lines from highly ionized Si, S, Ar,  and Ca, which indicate a thin thermal plasma origin. In addition, a strong Fe-K$\alpha$ line at 6.4 keV was found, which is indicative of a low-ionized Fe-rich plasma. Reach et al. (2006), using the Infrared Array Camera (IRAC) images at 3.6, 4.5, 5.8, and 8 $\mu$m from the GLIMPSE legacy science program of the Spitzer Space Telescope, detected an area of irregularly structured infrared emission on the western side of G344.7-0.1. 

With the unprecedented capabilities of the Chandra X-ray Observatory and XMM-Newton telescope, it is now possible to perform high-quality imaging and spectroscopy that are particularly well suited to the study of distant SNRs, such as G344.7-0.1. In the past decade, several new and well-known SNRs have been studied at hard X-ray energies using these instruments, with very interesting results (e.g. Senda et al. 2003; Bamba et al. 2003; Yamauchi et al. 2004; Combi et al. 2006; 2008). 

%%%%%%%%%%%%%%%%%%%%%%%%%%%%%%%%%%%fig-2%%%%%%%%%%%%%%
\begin{figure}
\centering
\includegraphics[width=8cm,angle=0]{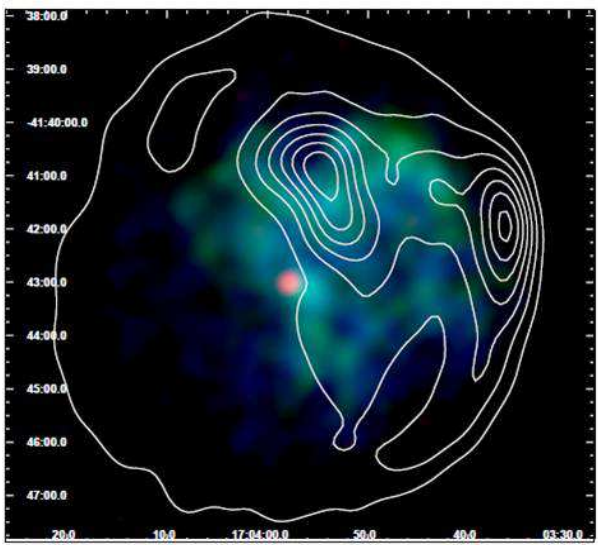}
\caption{Combined color-coded X-ray image of SNR G344.7-0.1 in the 0.5-8.0 keV energy band. Color coding is the same as for Fig.1. 
The position of the X-ray source CXOU J170357.8-414302, at the geometrical center of the SNR, is evident in the image (see text).}
\end{figure}
%%%%%%%%%%%%%%%%%%%%%%%%%%%%%%%%%%%%%%%%%%%%%%%%%%%%%

In this paper, we present an X-ray study of G344.7-0.1 and the point-like X-ray source discovered at the geometrical center of the SNR, using XMM-Newton and Chandra observations. Archival Spitzer data were also used to compare the radio, infrared, and X-ray morphologies of the SNR. The structure of the paper is as follows: in Sect. 2, we describe XMM-Newton and Chandra observations and the data reduction. X-ray analysis and results are presented in Sect. 3. The results from the Spitzer observations are presented in Sect. 4 and a search for radio, infrared, and optical counterparts are presented in Sect. 5. Finally, we discuss our results in Sect. 6 and present our conclusions in Sect. 7.

%------------------------------------------------------------------

\section{Observations and data reduction}

The field of G344.7-0.1 was observed by the Newton X-ray Multi-Mirror Mission (XMM) 
observatory with the European Photon Imaging Camera (EPIC) pn and MOS cameras. Furthermore, two Chandra 
X-ray observations conducted with the ACIS camera are available. Such a large set of observations provide us the possibility, 
for the first time, to perform a detailed X-ray analysis of SNR G344.7-0.1.

The XMM data were analyzed with the XMM Science Analysis System (SAS) version
9.0.0 and the latest calibrations. Chandra observation were calibrated using CIAO (version 4.1.2) and CALDB (version 3.2.2). 
To exclude strong background flares, which can affect the observations, we extracted 
light curves of photons above 10 keV from the entire field-of-view of the cameras, and excluded 
time intervals up to 3$\sigma$ to produce a GTI file. Detailed information about the observations and the instrumental characteristics is given in
Table\,1.

%%%%%%%%%%%%%%%%%%%%%%%%%%%%%%%%%%%%%%%%%%%%%%%%%%%%%%%
\begin{table*}
\caption{Table of observations}
\label{obs}\centering
\begin{center}
\begin{tabular}{l c c l c c c c}
\hline
Satellite& \multicolumn{2}{c}{Chandra}&& \multicolumn{3}{c}{XMM-Newton} \\ 
\cline{2-3} \cline{5-7}
Obs-Id     & 4651 & 5336 && 0111210101 & 0111210401 & 0506410101 \\ 
Date			&24/05/2006& 29/04/2006		&&15/09/2000& 28/08/2001&13/09/2007\\
Start Time	 [UTC]	&20:50:13   & 07:30:55		&&02:29:26   & 16:56:39	&03:49:25	\\
Camera    		&ACIS-235678& ACIS-235678	&&	PN	& PN / MOS1,2	& PN / MOS1,2	\\
Filter 		 &  $--$		&	$--$			&&MEDIUM	& MEDIUM &	MEDIUM	\\
Modes (read/data) &TIMED/VFAINT&TIMED/VFAINT&&PFWE	& PFWE	&	PFWE	\\
Offset 		 &  ccd7 - on axis       & ccd7 -on axis		&&on-axis& on-axis 	& off-axis ($\sim$9.9 ')\\
Exposure [ks] & 21.020   & 	6.345		&&7.132   & 4.609	& 24.286		\\
GTI	[ks]	& 20.754   & 	6.265		&&6.552	 & 4.402	& 15.990		\\
\hline
\end{tabular}
\end{center}
All observation were taken from their respective satellite database. Note: PFWE refer to the Prime Full Window Extended observation mode. Pointing of Chandra observation is RA:17:04:04.1 , DEC:-41:44:16.7.
\end{table*}
%%%%%%%%%%%%%%%%%%%%%%%%%%%%%%%%%%%%%%%%%%%%%%%%%%%%%%%%%

\section{X-ray study of G344.7-0.1}
\subsection{X-ray images}

Because of the high spatial resolution and sensitivity of the data set, we were able to examine the X--ray morphology of the supernova remnant in detail. In Fig. 1, we show narrow-band images generated in the energy ranges 0.5-1.2 keV, 1.2-2.5 keV, and 2.5-8.0 keV, with superimposed radio contours at 843 MHz (Whiteoak \& Green 1996). In the soft energy range (i.e $<$1.2 keV), only a point-like object is detected. Extended X-ray emission is quite prominent at medium (1.2-2.5 keV) and hard (2.5-8.0 keV) energies. As can be seen, the overall diffuse X-ray emission correlates well with the brightest radio regions and some hard X-ray emission extends towards parts of the SNR where weak or no radio emission is observed. 

We combined these three images into a single false color image. Figure 2 shows an ACIS image of 
G344.7-0.1, where the overall structure of the diffuse X-ray emission matches the brightest regions 
of the radio remnant, which appears concentrated toward the northwest part of the source. 
Furthermore, it seems that the hard X-ray emission is more extended than the emission at medium energies.
Total X-ray fluxes, for each energy range, soft (0.5-1.2 keV), 
medium (1.2-2.5 keV), and hard (2.5-8.0 keV) are $F_\mathrm{0.5-1.2 keV} =
7.2\times 10^{-15}\ \mathrm{erg \,s^{-1} cm^{-2}}$, $F_\mathrm{1.2-2.5 keV} =
4.7\times 10^{-12}\ \mathrm{erg \,s^{-1} cm^{-2}}$, $F_\mathrm{2.5-8.0 keV} = 5.03\times 10^{-12}\ \mathrm{erg \,s^{-1} cm^{-2}}$, 
which correspond to 1\%, 47\,\%, and 52\% of the total observed X-ray flux 
($F_\mathrm{0.5-8.0 keV} = 1.0\times 10^{-11}\ \mathrm{erg \,s^{-1} cm^{-2}}$), respectively. 

The sensitivity of the Chandra observations allowed us to detect, for the first time, a soft 
X-ray point-like source at the geometrical center of the radio structure of the SNR. This central X-ray 
source is absent in the radio map, but clearly detected in the Chandra data (see Fig. 3). We used PWDetect
detection code (Damiani et al. 1997a; Damiani et al. 1997b) to improve the quality of the detection. 
Since the central soft source is surrounded by non-uniform SNR emission, mostly over 1.5 keV,
the detection procedure was performed onto an event list file restricted to soft energies (i.e. 0.5-1.5 keV).
The source is located at $\alpha_{\rm J2000}$=17:03:57.852, $\delta_{\rm J2000}$=-41:43:02.03 (hereafter we refer to this object as CXOU J170357.8-414302). A total of 49 photons were detected in a single event file in the 0.5-1.2 keV. 
The X-ray source has a $\sim$17$\sigma$ significance above local background, being a quite robust detection. 

To check whether CXOU J170357.8-414302 is a point-like object or not, we searched for extended X-ray emission (i.e a pulsar wind nebula) that might be associated with the source. For this purpose, we applied the PWDetect detection code out to a radius of 16 arcsec. As a result, the analysis shows no trace of significant extended X-ray emission around the position of the object. In Fig. 3, we show the surroundings of CXOU J170357.8-414302. The source is clearly point-like.  

%%%%%%%%%%%%%%%%%%%%%%%%%%%%%%%%%%fig-3%%%%%%%%%%%%%%%%%%%%%%
\begin{figure}
\centering
\includegraphics[width=9cm,angle=0]{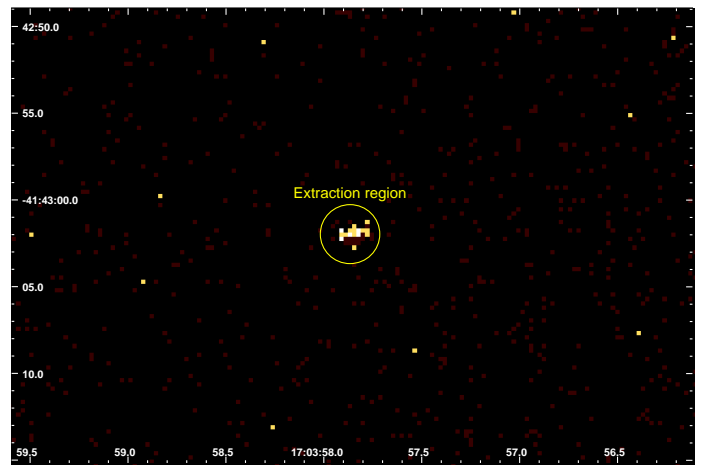}
\caption{Chandra ACIS image, with a size of 56"$\times$ 36", of CXOU J170357.8-414302 and its surroundings
in the 0.5-1.5 keV band. The pixel size is 0.246" (i.e., 1/2 of the original ACIS pixel size).
The yellow circle around the object shows the extraction region of 1.7" radius used for
the spectral analysis. No pulsar wind nebula is observed in the image.}
\end{figure}
%%%%%%%%%%%%%%%%%%%%%%%%%%%%%%%%%%%%%%%%%%%%%%%%%%%%%%%%%%%

\subsection{Spectral analysis}

\subsubsection{The SNR G344.7-0.1}

XMM-Newton and Chandra spectra were extracted for G344.7-0.1.
For the EPIC camera, we used {\sc evselect} SAS task with the appropriate parameters for 
PN and MOS 1/2 cameras. ACIS X--ray spectra was also extracted using the specific CIAO 
{\sc specextract} task for extended sources. The extraction radii used for the entire SNR are 
2.8 and 4.3 arcmin for Chandra and XMM data, respectively. We got nine EPIC (PN 3, 
MOS 4) and two ACIS X--ray spectra. Background spectra were also extracted from regions 
in which the SNR does not emit X--rays\footnote{We examined the effects of the background 
aperture size on the spectral fitting. The parameters obtained from EPIC and ACIS spectral 
fit are consistent, with differences within 1$\sigma$ uncertainties.}.

%%%%%%%%%%%%%%%%%%%%%%%%%%%%%%%%%fig-4%%%%%%%%%%%%%%%%%%
\begin{figure}
\centering
\includegraphics[width=6cm,angle=270]{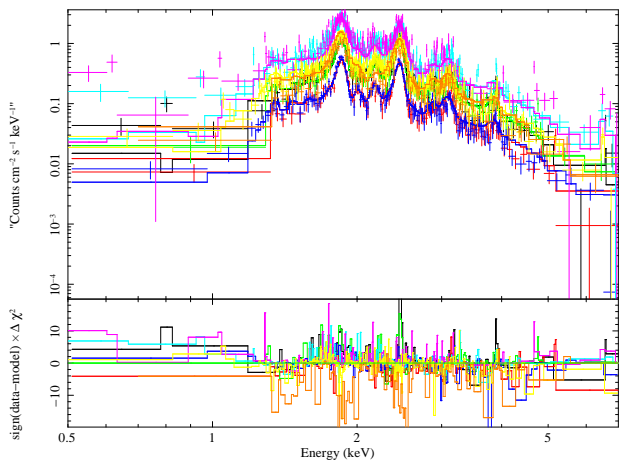}
\caption{{\bf Upper panel:} EPIC pn (in magenta, cyan and yellow colors) and MOS1/2
(in black, red, green and blue colors) spectra of the SNR G344.7-0.1, for the distinct
observations. ACIS-I X-ray spectra of the SNR is indicated in orange. Solid lines
indicate the best-fit ({\sc pshock}) model (see Table 2). {\bf Lower panel:} Chi-squared
residual of the best-fit model.}
\end{figure}
%%%%%%%%%%%%%%%%%%%%%%%%%%%%%%%%%%%%%%%%%%%%%%%%%%%%

Figure 4 shows the background-subtracted spectra obtained from the XMM-Newton and Chandra observations. 
We grouped the extracted EPIC and ACIS spectra with a 
minimum of 36 and 16 counts per spectral bin, respectively. The spectral analysis was performed 
using the XSPEC package (Arnaud, 1996). The overall X-ray spectrum of G344.7-0.1 exhibits several strong emission lines at the energies of 1.34, 1.85, 2.00, 2.18, 2.44, 2.86, 3.12, 3.87, and 6.4 keV. Following the interactive guide for 
ATOMDB\footnote{http://cxc.harvard.edu/atomdb/WebGUIDE/index.html}, 
we were able to identify the most prominent emission lines according to their transition intensities. Observed
features in the spectra correspond to atomic transitions of Mg XI, Si XIII, Si XIV, S XV, Ar XVII, Ca XIX, and Fe XXV.  
However, because of the moderate spectral resolution of the EPIC and ACIS cameras, most of them are strongly 
affected by line blending, biasing the identification and abundance determination of single 
chemical elements. 

X-rays from most SNRs come from a hot thin plasma consisting of ejecta and swept-up interstellar medium. The
X-ray spectrum of this hot plasma is usually fitted by a non-equilibrium ionization (NEI) model, with different abundances
for each chemical element. In this case, the X-ray spectrum was fitted with a PSHOCK model (Borkowski et al. 2001) modified by a low-energy absorption
model WABS (Morrison \& McCammon 1983). To evaluate systematics, we fitted XMM and Chandra data separately, and compared results with those obtained from a simultaneous EPIC+ACIS fit (see Table 2). We decided to avoid energy channels that are not well calibrated, ignoring energies below 0.5 keV and above 8.0 keV in the X--ray spectral analysis.

%%%%%%%%%%%%%%%%%%%%%%%%%%%nueva%%%%%%%%%%%%%
\begin{table*}
\caption{X--ray spectral parameters of the G344.7-0.1}
\label{spec}
\begin{center}
\begin{tabular}{l | l | l l l l l}
\hline
Parameters & \multicolumn{1}{c |}{Whole SNR}        & \multicolumn{1}{c}{Centre}&
\multicolumn{1}{c}{SE}& \multicolumn{1}{c}{Shock Front}& \multicolumn{1}{c}{Radio peak}&
\multicolumn{1}{c}{Hole} \\
\hline
{\bf WABS}                 &                                                &             
                   &                                        &
        &                                &                        \\
$N_{\sc H}$  [$\times$10$^{22}$] &4.91 ($\pm$0.02)&5.7 ($\pm$0.2) &5.3 ($\pm$0.1) &5.5
($\pm$0.2) & 4.3 ($\pm$0.3)&7.0($\pm$0.7)\\
\hline
{\bf PSHOCK}            &                  &               &               &             
&                                &                \\
kT [keV]                     & 1.17 ($\pm$0.02)  & 0.89 ($\pm$0.09) &1.81 ($\pm$0.2) 
&0.80($\pm$0.07) &0.83 ($\pm$0.09) &0.77 ($\pm$0.15)\\
abundance                   & 4.7 ($\pm$0.3)  &2.8 ($\pm$1.1) &4.4($\pm$0.9) 
&6.1($\pm$1.8) & 2.1($\pm$0.6)  & 9.5($\pm$0.9)        \\
$\tau_{\sc ul}$ [$\times$10$^{11}$] & 2.5 ($\pm$1.0)&1.4 ($\pm$0.9)&1.1 ($\pm$0.8)&3.5
($\pm$1.3)&3.0 ($\pm$0.9)&2.6 ($\pm$0.6) \\
Norm$\dag$[$\times$10$^{-3}$]& 33.2($\pm$0.2) & 4.4 ($\pm$1.5)  &2.4($\pm$0.9) 
&4.74($\pm$0.9)&5.8($\pm$1.4) &1.5 ($\pm$1.0)\\
E.M.  &7.80$\times$10$^{58}$& 1.03$\times$10$^{58}$&
5.69$\times$10$^{57}$&1.11$\times$10$^{58}$& 1.37$\times$10$^{58}$&3.54$\times$10$^{57}$\\
Flux [cgs]
&7.25$\times$10$^{-10}$&1.14$\times$10$^{-10}$&6.12$\times$10$^{-11}$&10.79$\times$10$^{-11}$&4.74$\times$10$^{-11}$&7.56$\times$10$^{-11}$\\
\hline
$\chi^2_\nu$/d.o.f.         &1.5 / 4037  &1.1 / 70  &1.4 / 245 &1.12 / 142 &1.1 / 101
&1.0 / 64\\
\hline
\end{tabular}
\end{center}
%%%%%%%%%%%%%%%%%%%%%%%%%%%%%%%%%%%%%%%%%%%%%%
$N_{\rm H}$ is in units of cm$^{-2}$. $\tau$ is in units of s cm$^{-3}$.
Fe-abund is  relative to solar values of Anders \& Grevesse (1989).
Normalization is defined as 10$^{-14}$/4$\pi$D$^2$$\times \int n_H\,n_e dV$, where D is
the distance in units of cm, $n_{\sc H}$ is the Hydrogen density [cm$^{-3}$], $n_e$ is
the electron density [cm$^{-3}$], and V is the volume [cm$^{3}$] in units of 10$^{-3}$.
$\tau_{\sc ll}$ lower limit was fix to zero while $\tau_{\sc ul}$ upper limits,
ionization time-scales was left as free parameter and is expressed in units of
s\,cm$^{-3}$. EM is the emission measure of the X--ray emitting plasma in units of cm$^3$.
Values in parentheses are single parameter 90\% confidence interval.
\end{table*}
%%%%%%%%%%%%%%%%%%%%%%%%%%%%%%%%%

In the fitting procedure, we initially froze the individual abundances to solar, and just varied the absorption 
($N_{\rm H}$), temperature ($kT$), ionization timescale upper limit ($\tau_{\sc ul}$), and model 
normalization. The ionization timescale lower limit ($\tau_{\sc ll}$) was fixed to zero, because the {\sc pshock} model describes 
a plasma with a linear distribution of timescales, ranging from the immediately post-shock region (where $\tau$=$\tau_{\sc ll}$) to 
the plasma at the largest distance from the shock front (at $\tau_{\sc ul}$). 
Once we obtained the main parameters of {\sc pshock}, we treated the abundance as a free parameter. 
We note that as previously reported by Yamauchi et al. (2005), the Fe line at 6.4 keV is likely 
also present in our spectra, although marginally detected. All results from our spectral analysis are shown in Table 2.

Finally, we compared the best-fit parameters obtained using a {\sc pshock} model, with those reported by Yamaguchi et al. (2005), obtained with the ASCA data, using a thin thermal plasma model (NEI) with an extra component for producing the Fe-K line. The higher quality photon statistics and spatial resolution of the XMM-Newton and Chandra data have obviously improved the quality of the spectral fit. The {\sc pshock} model naturally reproduces the observed spectrum (no extra component is required), and provides a good first approximation to the physical state of the plasma. Compared to the (NEI+Fe K line) model used by Yamaguchi et al. (2005), in the {\sc pshock} model every fitted parameter is more tightly constrained.

\subsubsection{The source CXOU J170357.8-414302}

%%%%%%%%%%%%%%%%%%%%%%%%%%%%%%%%fig-5%%%%%%%%%%%%%%%%%%%%5
\begin{figure}
\centering
\includegraphics[width=5.8cm,angle=270]{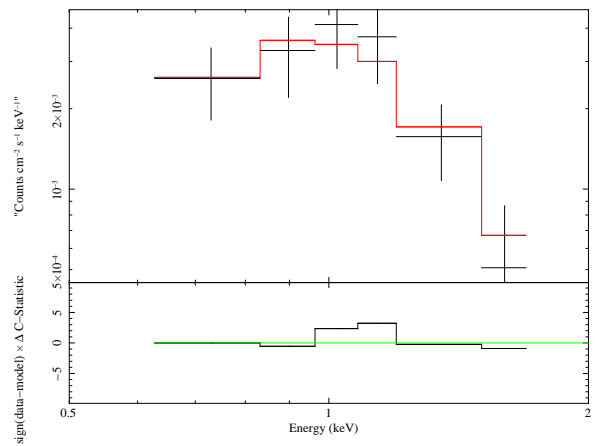}
\caption{Fit of the CXOU J170357.8-414302 spectrum extracted from the 1.7" aperture with an absorbed 
power-law model (see text).}
\end{figure} 
%%%%%%%%%%%%%%%%%%%%%%%%%%%%%%%%%%%%%%%%%%%%%%%%%%%%%%%%%%%

The point-like source CXOU J170357.8-414302, detected at the geometric center of the radio SNR, is marginally detected in the XMM-Newton images. There are three available XMM-Newton observations of the region. The object is located $\sim$ 2 arcmin offset from the aim-point and the extraction radius enclosing 90\% of the energy is 9 arcsec. Unfortunately, the source is at a position where the X-ray emission from the diffuse SNR gas is intense, thus veiling the intrinsic spectrum of the source. Out of the expected 60 photons from the PN spectrum, we obtained a total of 135 photons for the background corrected spectrum, i.e. more than 50\% of the photons come from the diffuse SNR X-ray emission. This situation is strongly improved in the Chandra observation.

Since the point-like source CXOU J170357.8-414302 is embedded in the diffuse 
X-ray emission of the SNR (see Fig.1), the source extraction region should 
avoid including SNR photons. Chandra data were used for the analysis because of 
its high spatial resolution (see Sect. 2). 
Following the analysis in Pavlov \& Luna (2009), we chose a 1.7" radius and
used the CIAO {\sc psextract} script to extract a spectrum. 
The 1.7" radius aperture contains 57 photons in the 0.2-3.0 keV range. 
This corresponds to a source count-rate of 1.8$\times$10$^{-3}$ count/s. 
Background was extracted from the same region used for the SNR spectral analysis. 
The spectrum shown in Fig. 5 is background corrected. 
We group the spectra with a minimum of 2 photons per bin because of the low count-rate.

The CXOU J170357.8-414302 spectrum was initially fitted by a single power-law (PL) model that yields 
an index $\Gamma$=9.5($\pm$2) and a normalization of 2.4($\pm$0.4)$\times$10$^{-4}$ cm$^{-2}$.
The absorption-corrected X--ray flux is $F_{\rm x}$=7.8$\times$10$^{-14} \mathrm{erg\, s^{-1} cm^{-2}}$ 
in the 0.7$-$2.0 keV band. The fit is acceptable in terms of the minimum $\chi^2$ ($\chi^2_\nu$ = 0.43 
for 41 d.o.f). We also fitted an absorbed thermal APEC model, which yields a 
neutral hydrogen absorption column $N_{\rm H}$=1.2($\pm$0.9)$\times$10$^{22}$ cm$^{-2}$
and a temperature $kT$=0.36$\pm$0.1 keV with sub-solar abundance of 0.66$\pm$0.8. In this case, the 
absorption-corrected X--ray flux is $F_{\rm x}$=6.9$\times$10$^{-14} \mathrm{erg\, s^{-1} cm^{-2}}$. The fit is acceptable in terms of the minimum $\chi^2$ ($\chi^2_\nu$ = 0.51 for 39 d.o.f).
 
Finally, to search for variability in our ACIS-I observation, 
we used the photon arrival times in the 0.5-2.0 keV band. With the frame readout 
time ($t_\mathrm{frame} = 3.24 \mathrm{s}$), we can search for variability
on timescales longer than  2\, $t_\mathrm{frame} = 6.48 \mathrm{s}$. No hints of variability were 
detected during the observation, indicating that the pulsar period, if any, should probably be shorter than 1 s.
The source has not been detected in the XMM-Newton data because of the high background of the 
EPIC camera. Therefore, no variability study could be performed on EPIC observations.

%%%%%%%%%%%%%%%%%%%%%%%%%%%%%%%fig-6%%%%%%%%%%%%%%%%%%%%%%
\begin{figure}
\centering
\includegraphics[width=8.5cm,angle=0]{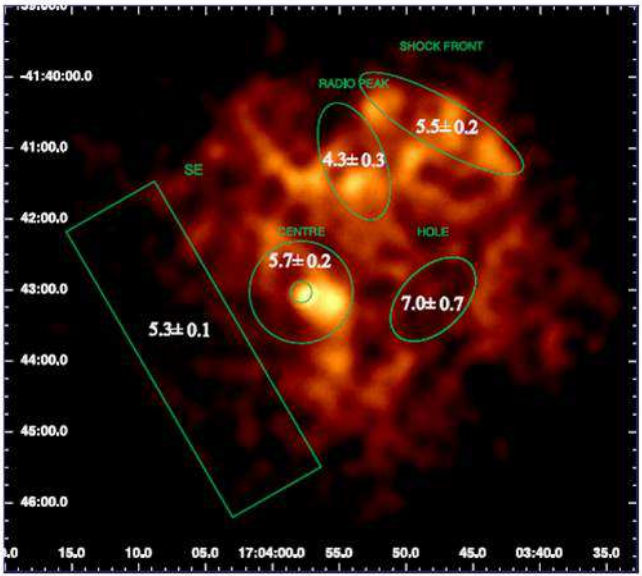}
\caption{Distribution of the column density $N_{\rm H}$ for different regions (i.e., centre, hole, radio peak, shock front and southeast:SE) expressed in units of 10$^{22}$ cm$^{-2}$. $N_{\rm H}$ was derived using the model summarized in Table 2. Note that the central point-like source was excluded  from the center region.}
\end{figure} 
%%%%%%%%%%%%%%%%%%%%%%%%%%%%%%%%%%%%%%%%%%%%%%%%%%%%%%%%%%

\subsubsection{Spatially resolved spectral analysis}

If CXOU J170357.8-414302 and the SNR were physically related, both sources should have similar neutral hydrogen absorption column. To check for possible $N_{\rm H}$ spatial variations across the SNR region (possibly due to an extremely inhomogeneous foreground medium), we extracted spatially resolved X-ray spectra at different regions (defined in Fig. 6) of the SNR. For this purpose, we used the same model to describe the global properties of the diffuse X-ray emission. As a result, we found that the neutral hydrogen absorption column of different regions have similar values to those obtained by analyzing the global spectrum of the remnant. The X-ray spectral parameters of different regions of G344.7-0.1 are shown in Table 3. This picture is consistent with the results obtained from the infrared analysis presented in Sect. 4.

As shown in Table 2, there are small variations in the temperature of the different regions, probably due to the efficiency of the heating processes, changes in the ISM density, or the shock wave interactions. However, the global temperature ($kT \sim$1.2 keV) agrees with that expected from  middle-aged SNRs (e.g. Williams \& Chu 2005). We also studied the spatial-abundance variations, which become higher (by a factor two or even three) at the region named {\sc front-shock}. It suggests that chemical inhomogeneities are detected in the ejecta. The other fitted parameters can be considered unchanged within the expected errors. Deep X--ray and radio observations are needed to determine changes in the spectrum along the different parts of the radio-emission, but this point is beyond the scope of this paper. 

\section{Infrared emission from G344.7-0.1}

Using Infrared Array Camera (IRAC) images at 3.6, 4.5, 5.8, and 8 $\mu$m from the GLIMPSE science program with the Spitzer Space Telescope, Reach et al. (2006) detected an area of irregularly structured infrared emission on the western part of G344.7-0.1. Although part of the infrared emission in the IRAC channels is coincident with the brightness regions of radio emission, no SNR structure was detected with IRAC by these authors, who suggest that the infrared colors of the structure observed with IRAC are compatible with ionized shocked gas and molecular shocks (see Fig. 22, Reach et al. 2006). 

It has been demonstrated that most SNR emit conspicuously at 24 $\mu$m (e.g. Borkowski et al. 2006; Williams et al. 2006; Morton et al. 2007). This emission traces warm dust (very small grains, VSG) stochastically heated to temperatures of around 30-100 K (Li \& Draine 2001). In SNRs, this dust component is heated by electrons and X-ray photons within the hot thermal gas (Dwek \&
Werner 1981; Hines et al. 2004). Therefore, a correlation between X-ray and 24 $\mu$m emission is expected.

To investigate the 24 $\mu$m emission from hot grains in G344.7-0.1, we used a Spitzer–-MIPS (Rieke et al. 2004) observation of
the region performed in October 2006 (AORs. \#20496896, \#20497152, and \#20497408). The MIPS basic calibrated
data (BCD) were downloaded from the Spitzer archive. These images were processed with the regular MIPS pipeline
(version S18.7.0), and then mosaicked using MOPEX (version 18.3.1) and the standard MIPS 24 $\mu$m mosaic pipeline. In
Fig.7, left, central and right panels we show the MIPS image with superimposed radio contours, without the radio contours, and a composite of the infrared, optical, and X-ray images, respectively. In contrast to what is observed in the IRAC bands by Reach et al. (2006), the 24 $\mu$m emission fills the radio contours and exhibits a good correlation with the diffuse X-ray emission. 

In general, the 24 $\mu$m emission exhibits a filamentary morphology that is strongly correlated with regions of radio emission. 
In addition, two external infrared filaments seen at 24 $\mu$m are not observed in X-rays. They coincide in location with
the southwest and northeast filaments observed at radio wavelengths. In Fig.7 (right panel), we show an RGB color-coded
image of the SNR. We used red for the 24 $\mu$m emission, green for the IRAC 8 $\mu$m emission (AOR \#11957248), and blue for an
DSS optical image in the R band. Emission from interstellar medium at 8 $\mu$m is dominated by polycyclic aromatic hydrocarbon
(PAH) bands.

%%%%%%%%%%%%%%%%%%%%%%%%%%%%%%%fig-7%%%%%%%%%%%%%%%%%%%%%%
\begin{figure*}
\centering
\includegraphics[width=18.0cm,angle=0]{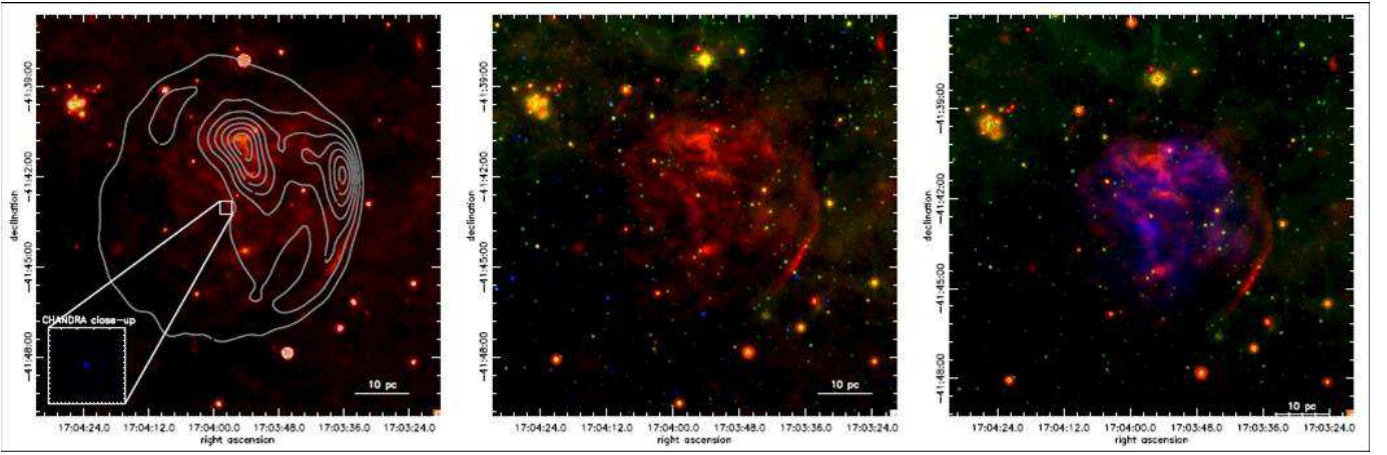}
\caption{{\bf Left panel:} MIPS 24 $\mu$m mosaic of G344.7-0.1. Radio contours are overplotted. The square marks the position of the punctual X-ray source observed with Chandra (small window in the left-bottom corner). Scale was determined assuming a distance of 14 kpc for the SNR. {\bf Central  panel:} Three-color image of G344.7-0.1 (red is MIPS 24 $\mu$m, green is IRAC 8 $\mu$m, and blue is optical R band). {\bf Right panel:} RGB composition of MIPS (24 $\mu$m; red), IRAC (8 $\mu$m; green), and Chandra (0.5–8.0 keV energy band; blue).}
\end{figure*} 
%%%%%%%%%%%%%%%%%%%%%%%%%%%%%%%%%%%%%%%%%%%%%%%%%%%%%%%%%%}

%%%%%%%%%%%%%%%%%%%%%%%%%%%%%%%%%%%%%%%%%%%%%%%%%%%%%%%%%%}

\section{Search for radio, infrared, and optical counterparts to CXOU J170357.8-414302}

To identify possible radio, infrared or optical counterparts 
within the location error box of the central X-ray source, we used VLA data, inspected the 2 Micron All Sky Survey 
(2MASS, Cutri et al. 2003), and searched the USNO B1.0 optical catalog (Monet et al. 2003). 

To find evidence of radio emission from the CXOU J170357.8-414302 source, we 
explored the radio data in the National Radio Astronomy Observatory
(NRAO) archive. The region has been barely observed and therefore data from 
only three projects were retrieved. Among them only the AD260,
conducted on 1991 February 23 with the Very Large Array (VLA), is
available in continuum. This observation was carried out in
CD-configuration at 20 cm wavelength, amounting to a total on-source
time of 40 minutes. The instrumental setup included two intermediate
frequency (IF) pairs with 50 MHz bandwidth each. The data set was
processed using the AIPS software package of NRAO following the
standard procedures for continuum calibration of interferometers.
The flux density scale was set using the primary amplitude
calibrator 3C286, whereas the phase calibration was performed by
repeated observations of the nearby phase calibrator 1730-130. At
the end of a self-calibration deconvolution iteration scheme, we
produced several versions of the radio map using different values
for the ROBUST and UVRANGE parameters to enhance the
point-like sources. A close inspection indicates no source detection
inside or in the vicinity of the Chandra error circle. The resulting
radio flux upper limit (3-$\sigma$) at 20 cm is estimated to be 0.9
mJy/beam.

At the infrared and optical part of the spectrum, we found the sources USNO B1.0 0482-0503281 and 
2MASS J17035785-4143020 to be consistent with the Chandra position of the putative point-like X-ray source. 
Figure 8 displays the probable USNO R-band counterpart of CXOU J170357.8-414302. 
Table 3 displays the photometric properties of the infrared 
and optical counterparts of the central source.

%%%%%%%%%%%%%%%%%%%%%%%%%%%%%%%%%%%%%%%%%%%%%%%%%%%%%%%
\begin{table}
\caption{Photometric data}
\label{fotom}
%\begin{center}
\begin{tabular}{c c c c c}
\hline
Filter & Apparent & A$_{V}$ & De-reddened & Flux\\
       & magnitude &        & magnitude & erg s$^{-1}$ cm$^{-2}$ \AA$^{-1}$\\
\hline
B     & 15.37$^{a}$ & 7.4 &  7.97 & (0.429 $\pm$ 0.004)10$^{-11}$ \\
R     & 14.22$^{a}$ & 4.2 & 10.04 & (0.167 $\pm$ 0.002)10$^{-12}$ \\
I     & 13.66$^{a}$ & 2.7 & 10.97 & (0.342 $\pm$ 0.003)10$^{-13}$ \\
J     & 12.77$^{b}$ & 1.6 & 11.19 & (0.110 $\pm$ 0.001)10$^{-13}$ \\
H     & 12.32$^{b}$ & 1.0 & 11.34 & (0.374 $\pm$ 0.003)10$^{-14}$ \\
K$_s$ & 12.24$^{b}$ & 0.6 & 11.61 & (0.900 $\pm$ 0.008)10$^{-15}$ \\
\hline
\end{tabular}
%\end{center}
\\
$^{a}$ USNO-B1 (United States Naval Observatory B1.0 Catalog)\\
$^{b}$ 2MASS All-Sky Point Source Catalog (PSC)\\
\end{table}

\section{Discussion}

\subsection{The origin of the diffuse X-ray emission in G344.7-0.1}

The X-ray analysis presented above has detected centrally bright X-ray emission from SNR G344.7-0.1 that is apparently thermal.
Its spectrum is dominated by prominent atomic emission lines such as Mg XII, Si XIII, Si XIV, S XV, Ar XVII, Ca XIX, and Fe XXV, and is represented well by a plane-parallel shock plasma model with variable abundances. The most intense region of radio emission and the diffuse X-ray emission of G344.7-0.1 correlate spatially well with an area of irregularly structured infrared emission detected with the Spitzer Space Telescope (see fig. 7 in  Reach et al. 2006). These results indicate that the irregular morphology of the remnant is caused by the expansion through a dense interstellar medium with a density gradient towards the west of the remnant.

It is widely accepted that the soft X-ray emission from middle-aged SNRs is produced by their expansion against a dense medium,
while the hard component originates in the interior heated by a fast shock in the early stage of the SNR evolution. 
In the case that we have studied, the soft X-ray emission arises mainly from the compact central source CXOU J170357.8-414302, and the thermal medium/hard X-ray emission, which correlates with regions of enhancement radio and infrared emission along the western edge of the SNR, seems to be associated with hot gas in the remnant interior, heated by the passage of the shock wave.

On the basis of information gathered at radio and X-ray wavelengths, it has been possible
to delineate the evolution of G344.7-0.1. We first estimated the volume $V$ of the X-ray emitting plasma. Assuming that the plasma fills
a sphere with a diameter of 6 arcmin, we obtained a volume of 4.9$\times$ 10$^{58}$ cm$^{3}$ at a mean distance of 14 kpc. Using the emission
measure (EM) determined from the spectral fitting, we estimated the electron density of the
plasma $n_{e}$=$\sqrt{EM/V}$ to be 1.26 cm$^{-3}$. In this case,
the density of the nucleons was simply assumed to be
the same as that of electrons. The age $t$ was then determined using the upper limit ionization timescale, 
$\tau_{ul}$, by $t$= $\tau_{ul}$/$n_{e}$. As a result, the elapsed time after the plasma was heated is $\sim$ 6$\times$ 10$^{3}$ yr. 
The total mass of the plasma $M_{\rm total}$ was estimated to be $M_{total}$=$n_{e}$$V$$m_{\rm H}$ $\sim$ 54$M_{\odot}$,
where $m_{\rm H}$ is the mass of a hydrogen atom.

On the other hand, we can independently estimate the age of the SNR using standard arguments based on the Sedov dynamics (Sedov 1959).
Assuming that the SNR expansion is adiabatic and adopting a mean distance of 14 kpc for G344.7-0.1, we found that the SNR radius should be $R\sim$ 16 pc. If the SNR expansion occurs in a medium with density $n \sim$ 0.2 cm$^{-3}$, and the released SN total energy is $E \sim$ 1.0$\times$10$^{51}$ erg (Spitzer  1998), then the SNR age is $t \sim$ 6.5$\times$ 10$^{3}$ yr. This value is consistent with that obtained from the X-ray emitting plasma.

%%%%%%%%%%%%%%%%%%%%%%%%%%%%%%%fig-8%%%%%%%%%%%%%%%%%%%%%%
\begin{figure}
\centering
\includegraphics[width=7.5cm,angle=0]{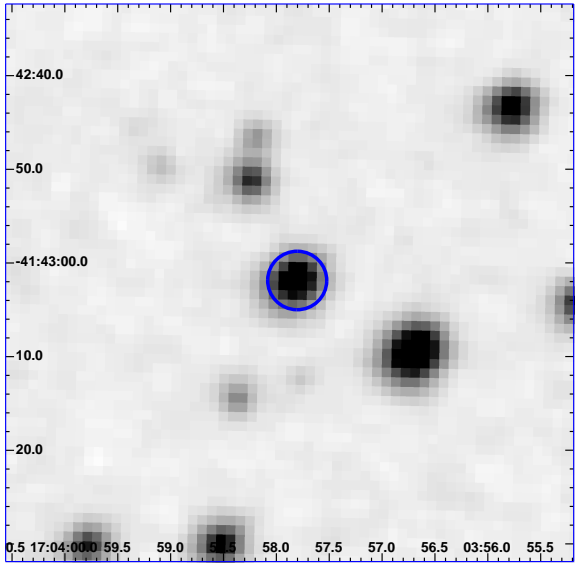}
\caption{USNO R-band counterpart of the central X-ray source CXOU J170357.8-414302. The position error circle (in blue) is
shown in the image.}
\end{figure} 
%%%%%%%%%%%%%%%%%%%%%%%%%%%%%%%%%%%%%%%%%%%%%%%%%%%%%%%%%%

\subsection{The nature of CXOU J170357.8-414302}

The point-like X-ray source located at the geometrical center of the SNR, exhibits steady X-ray flux emission that appears to exclude an accreting binary origin and soft thermal spectra that rules out a background active nucleus, lacks a radio counterpart, and contains no surrounding pulsar wind nebula. Therefore, the source displays some characteristics in common with the so-called CCO (see Pavlov et al. 2004, for a review), a new population of isolated neutron stars (NSs) with clear differences from isolated rotation-powered pulsars, and accretion-powered X-ray pulsars in close binary systems.

At present, only seven confirmed examples of CCOs are known (Gotthelf et al. 2005, Halpern \& Gotthelf 2009). The nature of these CCOs is still unclear. It is thought that the X-ray emission from CCOs is generally caused by the thermal cooling of the NS (e.g., Zavlin, Trumper \& Pavlov
1999), with typical temperatures of a few 10$^{6}$ K, as inferred from their thermal-like spectra. They have X-ray luminosities ($L_{\rm X}$) in
the range of 10$^{33}$-10$^{34}$ erg s$^{-1}$ and display X-ray spectra characterized by a black-body model with temperatures ($kT$) in the
range of 0.2-0.5 keV or a power-law model with very steep index $\Gamma$ (see Pavlov et al. 2003). Halpern \& Gotthelf (2009) suggested that these objects could be weakly magnetized NSs ($B\sim10^{10}$ G), i.e., a kind of ``anti-magnetars''. 

To verify the other physical parameters of CXOU J170357.8-414302, we computed its $L_{\rm X}$ and spin-down luminosity $\dot{E}$, to compare with well-known CCOs objects (Pavlov et al.  2003). Adopting again a mean distance of 14 kpc, we determined a total unabsorbed X-ray flux of  $F_{0.7-2.0}$=6.9$\times$10$^{-14}$ ergs cm$^{-2}$ s$^{-1}$, which corresponds to an unabsorbed luminosity $L_{\rm X}$= 1.6$\times$10$^{33}$ ergs s$^{-1}$. A rough estimate of the spin-down luminosity can be derived using the empirical formula by Seward \& Wang (1988), log $L_{\rm X}$ (ergs s$^{-1}$)= 1.39 log $\dot{E}$ - 16.6, which implies that $\dot{E}$= 6.7$\times$10$^{35}$ ergs s$^{-1}$. This value of $L_{\rm X}$ agrees with those in the range listed by Pavlov et al. (2004) and Halpern \& Gotthelf (2009) for CCO objects. The second quantity, $\dot{E}$, falls below the empirical threshold for generating bright wind nebulae of $\dot{E_{c}}$ $\approx$ 4$\times$10$^{36}$ ergs s$^{-1}$. All the results obtained are typical of CCO found in other supernova remnants (e.g. Pavlov et al. 2004; Gotthelf et al. 2008).

Nevertheless, we found an infrared and optical source that is positionally coincident with the point-like X-ray source, whose 
near-infrared colors ($J-H \simeq 0.08$ and $H-K \simeq 0.45$) are typical of an early-K star (giant or dwarf). If this source were the counterpart of the X-ray source, then the X-ray-to-optical flux ratio would be $F_{\rm X}$/$F_{\rm H}$ $\sim$ 7.9$\times$10$^{-3}$, which may exclude a neutron star origin of the central source (Fesen et al. 2006). However, the stellar density in the direction of G344.7-0.1 ($l = 344.7$, $b = -0.2$ deg) is very high. The 2MASS catalog contains 4638 point-like sources in a radius of 3 arcmin from the center of G344.7-0.1, 1318 of them with $J \le 14.5$ mag. Thus, a chance near-alignment between a foreground star and the central object found inside the SNR seems probable. 

To evaluate probability of the chance alignment between the central X-ray object and the central position of G344.7-0.1, we computed the probability that the distance between the geometrical center of the SNR and the position of the point-like source is smaller than the observed one. This is expressed by

\begin{equation}
P(d < d_{\rm obs}) = 1 - e^{-n\,\pi\,d^2} \approx n\,\pi\,d^2,
\end{equation}

\noindent where $n$ is the density of X-ray sources (in arcsec$^{-2}$), inside the radio SNR contours ($n \sim 0.00478$ arcsec$^{-2}$). The distance from the X-ray source to the SNR center exhibits an offset of d$\sim$14.2". We obtained a probability $P(d < d_\mathrm{obs})$ =  0.0035, i.e. just a 0.35\% chance alignment probability.

We also calculated the chance association of the observed 2MASS source with the CCO candidate. The total number of IR sources in the SNR is 4670, thus the source density $n$ is 0.0083325. Therefore, according to the previous relation, the probability that the X-ray to 2MASS PSFs are mismatched at a distance $d$ is $P(\sigma_{\rm X} - \sigma_{\rm spitzer})=0.16$, i.e., a chance association probability of $\sim$16\%.

Although no robust value of the distance to CXOU\,J170357.8-414302 can be determined, 
one may attempt to determine it from its visible and near-infrared colours. Assuming a 
spectral type K0 for a dwarf star, the source may be placed at a distance of between 
450 and 800 pc. If the star were a giant, the distance would increase to 
10 kpc, still far from the location of the SNR (at $\sim 14$ kpc). Nevertheless, we note
that only giants in close binary systems emit in X-rays.

If the infrared and optical sources were physically unrelated to the X-ray source, then CXOU\,J170357.8-414302 might be a CCO. However, this latter possibility seems improbable since the $N_{\rm H}$ value of the object is inconsistent with that for G344.7-0.1. If the point-like object and the SNR are physically associated then the foreground medium (between the observer and the source) must be extremely inhomogeneous. In this case, it may have been produced by a foreground molecular cloud with a density of 1000 particles/cm$^{3}$ extending 10 pc along the line of sight. This enormous cloud should be "in front" of the northwest region, but not in front of the "center" region, where a kind of "hole" in the interstellar absorption must be invoked. This dense and huge molecular cloud, if present, should be detectable by carbon monoxide (CO) observations. Unfortunately, at present  high-resolution CO observations of this region do not exist.

\section{Conclusions}

We have presented an X-ray study of the SNR G344.7-0.1, using
new XMM-Newton and Chandra observations. In addition, Spitzer–-MIPS observations
at 24 $\mu$m have been used to study the infrared morphology of the source. A clear correlation between 
X-ray and infrared emission indicates that radiation in both wavebands originated in the SNR. 
The IR flux has allowed us to characterize the medium in which the remnant is expanding. The detected diffuse X-ray emission 
correlates well with the brightest radio regions of the SNR. At both radio
and X-ray wavelengths, the western half of the remnant is substantially
brighter than the eastern half, where irregular infrared emission is also 
concentrated. The X-ray spectrum exhibits emission lines from Mg, Si, S,
Ar, and marginally Fe. Lines of Mg XI (1.34 keV), Si XIV (2.0 keV), and S XV (2.86
keV) were detected for the first time in this object. The X-ray characteristics 
suggest that the radiation has a thin thermal plasma origin, which is represented well 
by a plane-parallel shock plasma model. 
This provides a good first approximation to the physical state of the plasma.
The overall imaging and spectral properties of G344.7-0.1 favor the
interpretation of a middle-aged SNR ($\sim$ 6$\times$10$^{3}$ yr old), propagating
in a dense medium and likely encountering a molecular cloud on its west 
side. G344.7-0.1 provides an excellent laboratory to study the evolution and
interaction of a SNR with its surrounding medium.

In addition, we have reported the discovery of a soft point-like object at the geometrical
center of the G344.7-0.1 radio structure with some CCO characteristics. A broadband study from radio to the X-ray domain has shown that it might have  infrared and optical counterparts. If the infrared and optical sources were physically unrelated to the X-ray source, then CXOU\,J170357.8-414302 might be a CCO. However, this latter possibility seems unlikely since the $N_{\rm H}$ value of the object is inconsistent with that of G344.7-0.1. 

In summary, taking into account all gathered information we cannot confirm the nature of the object, and therefore its origin remains uncertain. Optical spectroscopic observations of the infrared/optical source are necessary to fix the distance. New Chandra observations, with a longer exposure time, will be important to improve our knowledge of this source.

\begin{acknowledgements}

We are grateful to the referee for his valuable suggestions and comments which helped us to improve the paper. The authors acknowledge support by DGI of the Spanish Ministerio de Educaci\'on y Ciencia under grants AYA2007-68034-C03-02/-01, FEDER funds, Plan Andaluz de Investigaci\'on Desarrollo e Innovaci\'on (PAIDI) of Junta de Andaluc\'{\i}a as research group FQM-322 and the excellence fund FQM-5418. J.A.C., J.F.A.C. and G.E.R are researchers of CONICET. J.F.A.C was supported by grant PICT 2007-02177 (SecyT). G.E.R. and J.A.C were supported by grant PICT 07-00848 BID 1728/OC-AR (ANPCyT) and PIP 2010-0078 (CONICET). J.L.S. acknowledges support by the Spanish Ministerio de Innovaci\'on y Tecnolog\'ia under grant AYA2008-06423-C03-03. 
PGP-G acknowledges support from the Ram\'on y Cajal Program, financed by the Spanish Government and/or the European Union. The National Radio Astronomy Observatory is a facility of the National Science Foundation operated under cooperative agreement by Associated Universities, Inc.

\end{acknowledgements}


\begin{thebibliography}{}

\bibitem[Anders 
\& Grevesse(1989)]{1989GeCoA..53..197A} Anders, E., \& Grevesse, N.\ 1989, \gca, 53, 197 

\bibitem[Arnaud(1996)]{1996ASPC..101...17A} Arnaud, K.~A.\ 1996, 
Astronomical Data Analysis Software and Systems V, 101, 17 

\bibitem[Bamba et al.(2003)]{2003ApJ...589..253B} Bamba, A., Ueno, M., 
Koyama, K., \& Yamauchi, S.\ 2003, \apj, 589, 253 

\bibitem[Borkowski et al.(2001)]{2001ApJ...548..820B} Borkowski, K.~J., 
Lyerly, W.~J., \& Reynolds, S.~P.\ 2001, \apj, 548, 820 

\bibitem[Borkowski et al.(2006)]{2006ApJ...642L.141B} Borkowski, K.~J., et 
al.\ 2006, \apjl, 642, L141 

\bibitem[Caswell et al.(1975)]{1975AuJPA..37....1C} Caswell, J.~L., Clark, 
D.~H., Crawford, D.~F., \& Green, A.~J.\ 1975, Australian Journal of Physics Astrophysical Supplement, 37, 1 

\bibitem[Combi et al.(2006)]{2006ApJ...653L..41C} Combi, J.~A., Albacete 
Colombo, J.~F., Romero, G.~E., \& Benaglia, P.\ 2006, \apjl, 653, L41 

\bibitem[Combi et 
al.(2008)]{2008A&A...488L..25C} Combi, J.~A., Albacete-Colombo, J.~F., \& Mart{\'{\i}}, J.\ 2008, \aap, 488, L25

\bibitem[Cutri et al.(2003)]{2003...} Cutri, R. M., Skrutskie, M. F., van Dyk, S., et al. 2003, VizieR Online
Data Catalog, II/246 (http://cdsweb.u-strasbg.fr/viz-bin/Cat?II/246)

\bibitem[Damiani et al.(1997a)]{1997ApJ...483..350D} Damiani, F., Maggio, 
A., Micela, G., \& Sciortino, S.\ 1997a, \apj, 483, 350 

\bibitem[Damiani et al.(1997b)]{1997ApJ...483..370D} Damiani, F., Maggio, 
A., Micela, G., \& Sciortino, S.\ 1997b, \apj, 483, 370

\bibitem[Dubner et al.(1993)]{1993AJ....105.2251D} Dubner, G.~M., Moffett, 
D.~A., Goss, W.~M., \& Winkler, P.~F.\ 1993, \aj, 105, 2251 

\bibitem[Dwek 
\& Werner(1981)]{1981ApJ...248..138D} Dwek, E., \& Werner, M.~W.\ 1981, \apj, 248, 138 

\bibitem[Halpern \& Gotthelf(2009)]{2009arXiv0911.0093H} Halpern, J.P., \& Gotthelf, E.V.\ 2009, \apj, submitted [arXiv0911.0093]

\bibitem[Hines et al.(2004)]{2004ApJS..154..290H} Hines, D.~C., et al.\ 
2004, \apjs, 154, 290 

\bibitem[Huang 
\& Thaddeus(1985)]{1985ApJ...295L..13H} Huang, Y.-L., \& Thaddeus, P.\ 1985, \apjl, 295, L13 

\bibitem[Fesen et al.(2006)]{2006ApJ...636..848F} Fesen, R.~A., Pavlov, 
G.~G., \& Sanwal, D.\ 2006, \apj, 636, 848 

\bibitem[Gotthelf et al.(2005)]{2005ApJ...627..390G} Gotthelf, E.~V., 
Halpern, J.~P., \& Seward, F.~D.\ 2005, \apj, 627, 390 

\bibitem[Gotthelf 
\& Halpern(2008)]{2008AIPC..983..320G} Gotthelf, E.~V., \& Halpern, J.~P.\ 2008, 40 Years of Pulsars: Millisecond Pulsars, Magnetars and More, 983, 320 

\bibitem[Li 
\& Draine(2001)]{2001ApJ...554..778L} Li, A., \& Draine, B.~T.\ 2001, \apj, 554, 778 

\bibitem[Morrison 
\& McCammon(1983)]{1983ApJ...270..119M} Morrison, R., \& McCammon, D.\ 1983, \apj, 270, 119

\bibitem[Morton et al.(2007)]{2007ApJ...667..219M} Morton, T.~D., Slane, 
P., Borkowski, K.~J., Reynolds, S.~P., Helfand, D.~J., Gaensler, B.~M., 
\& Hughes, J.~P.\ 2007, \apj, 667, 219 

\bibitem[Pavlov et al.(2003)]{2003ApJ...591.1157P} Pavlov, G.~G., Teter, 
M.~A., Kargaltsev, O., \& Sanwal, D.\ 2003, \apj, 591, 1157 

\bibitem[Pavlov et al.(2004)]{2004IAUS..218..239P} Pavlov, G.~G., Sanwal, 
D., \& Teter, M.~A.\ 2004, Young Neutron Stars and Their Environments, 218, 239 

\bibitem[Pavlov 
\& Luna(2009)]{2009ApJ...703..910P} Pavlov, G.~G., \& Luna, G.~J.~M.\ 2009, \apj, 703, 910 

\bibitem[Reach et al.(2006)]{2006AJ....131.1479R} Reach, W.~T., et al.\ 
2006, \aj, 131, 1479

\bibitem[Rieke et al.(2004)]{2004ApJS..154...25R} Rieke, G.~H., et al.\ 
2004, \apjs, 154, 25 

\bibitem[Sedov(1959)]{1959sdmm.book.....S} Sedov, L.~I.\ 1959, Similarity 
and Dimensional Methods in Mechanics, New York: Academic Press, 1959,  

\bibitem[Senda et al.(2003)]{2003ANS...324..151S} Senda, A., Murakami, H., 
\& Koyama, K.\ 2003, Astronomische Nachrichten Supplement, 324, 151

\bibitem[Seward 
\& Wang(1988)]{1988ApJ...332..199S} Seward, F.~D., \& Wang, Z.-R.\ 1988, \apj, 332, 199 

\bibitem[Spitzer(1998)]{1998ppim.book.....S} Spitzer, L.\ 1998, Physical 
Processes in the Interstellar Medium, by Lyman Spitzer, pp.~335.~ISBN 
0-471-29335-0.~Wiley-VCH , May 1998.

\bibitem[Sugizaki et al.(2001)]{2001ApJS..134...77S} Sugizaki, M., Mitsuda, 
K., Kaneda, H., Matsuzaki, K., Yamauchi, S., 
\& Koyama, K.\ 2001, \apjs, 134, 77

\bibitem[Whiteoak 
\& Green(1996)]{1996A&AS..118..329W} Whiteoak, J.~B.~Z., \& Green, A.~J.\ 1996, \aaps, 118, 329

\bibitem[Williams 
\& Chu(2005)]{2005ApJ...635.1077W} Williams, R.~M., \& Chu, Y.-H.\ 2005, \apj, 635, 1077 

\bibitem[Williams et al.(2006)]{2006ApJ...652L..33W} Williams, B.~J., et 
al.\ 2006, \apjl, 652, L33 

\bibitem[Yamaguchi et al.(2004)]{2004PASJ...56.1059Y} Yamaguchi, H., Ueno, 
M., Koyama, K., Bamba, A., \& Yamauchi, S.\ 2004, \pasj, 56, 1059

\bibitem[Yamauchi et al.(2005)]{2005PASJ...57..459Y} Yamauchi, S., Ueno, 
M., Koyama, K., \& Bamba, A.\ 2005, \pasj, 57, 459 

\bibitem[Yu et al.(2009)]{2009ApJ...690..440Y} Yu, Y.~S., Nordon, R., 
Kastner, J.~H., Houck, J., Behar, E., \& Soker, N.\ 2009, \apj, 690, 440

\bibitem[Zavlin et al.(1999)]{1999ApJ...525..959Z} Zavlin, V.~E., 
Tr{\"u}mper, J., \& Pavlov, G.~G.\ 1999, \apj, 525, 959

  
\end{thebibliography}
\end{document}